\pdfoutput=1
\documentclass[preprint,12pt]{elsarticle}
\usepackage[left=25mm, bottom=25mm, right=25mm, top=25mm]{geometry}

\usepackage{lineno,hyperref}

\usepackage{tabularx}
\makeatletter
\def\ps@pprintTitle{%
 \let\@oddhead\@empty
 \let\@evenhead\@empty
 \def\@oddfoot{}%
 \let\@evenfoot\@oddfoot}
\makeatother

\usepackage{amsmath}
\usepackage{amssymb}

\usepackage{float}

\biboptions{authoryear}

\begin{document}

\begin{frontmatter}

\title{Cloud removal in remote sensing images using generative adversarial networks and SAR-to-optical image translation}


\author[mymainaddress]{Faramarz Naderi Darbaghshahi}
\ead{f_naderi97@comp.iust.ac.ir}

\author[mymainaddress]{Mohammad Reza Mohammadi}
\ead{mrmohammadi@iust.ac.ir}
\author[mymainaddress]{Mohsen Soryani
\corref{mycorrespondingauthor}}
\cortext[mycorrespondingauthor]{Corresponding author}
\ead{soryani@iust.ac.ir}
\address[mymainaddress]{School of Computer Engineering, Iran University of Science and Technology \\University Road, Hengam Street, Resalat Square, Narmak, Tehran, P.O. Box 16846-13114, Iran}


\begin{abstract}
Satellite images are often contaminated by clouds. Cloud removal has received much attention due to the wide range of satellite image applications. As the clouds thicken, the process of removing the clouds becomes more challenging. In such cases, using auxiliary images such as near-infrared or synthetic aperture radar (SAR) for reconstructing is common. In this study, we attempt to solve the problem using two generative adversarial networks~(GANs). The first translates SAR images into optical images, and the second removes clouds using the translated images of prior GAN. Also, we propose dilated residual inception blocks~(DRIBs) instead of vanilla U-net in the generator networks and use structural similarity index measure (SSIM) in addition to the L1 Loss function. Reducing the number of downsamplings and expanding receptive fields by dilated convolutions increase the quality of output images. We used the SEN1-2 dataset to train and test both GANs, and we made cloudy images by adding synthetic clouds to optical images. The restored images are evaluated with PSNR and SSIM. We compare the proposed method with state-of-the-art deep learning models and achieve more accurate results in both SAR-to-optical translation and cloud removal parts.
\newline 
\end{abstract}


\end{frontmatter}



\section{Introduction}

Remote sensing plays a vital role in acquiring information for monitoring various fields such as recognizing buildings \citep{vakalopoulou2015building}, crop mapping \citep{kussul2017deep}, and land cover change detection \citep{lyu2016learning}. A study was conducted by Moderate Resolution Imaging Spectroradiometer (MODIS) over 12 years of continuous observations from Terra and over nine years from Aqua showed that clouds cover around 67\% of the Earth’s surface \citep{king2013spatial}. Therefore, cloud removal methods are considered to address this issue in optical remote sensing.

There are four categories for reconstructing the missing information on remote sensing data \citep{shen2015missing}: (1) spatial-based methods; (2) spectral-based methods; (3) temporal-based methods; and (4) hybrid methods. Since clouds cause missing information, we can use these methods for cloud removal.

Spatial-based methods restore corrupted pixels using cloud-free regions. \cite{maalouf2009bandelet} use bandelet transform and the multiscale geometrical grouping. \cite{meng2017sparse} present an adaptive patch method based on a sparse dictionary learning algorithm. These methods usually fail to recover contaminated pixels if cloud regions are large \citep{shen2015missing}.

Spectral-based methods use different sensors to acquire additional data for reconstructing missing information. The auxiliary data can be obtained by optical sensors \citep{zhang2018cloud}, but they also can be contaminated by clouds. To solve this problem, additional sources must be able to pass-through clouds, e.g., near-infrared \citep{enomoto2017filmy,wang2005new} or SAR \citep{gao2020cloud,li2019thick,meraner2020cloud}.

Temporal-based methods estimate missing information by integrating cloud-free correspondence images at different times. \cite{chen2019thick} present an end-to-end deep learning architecture to detect and remove clouds from high-resolution satellite data. \cite{xu2016cloud} remove clouds with combined coefficients from the reference image (cloud-free areas) and the dictionary learned from the target image (cloudy areas). Requiring cloud-free images with short intervals is the limitation of these methods.

Hybrid methods utilize a combination of the three previous methods for cloud removal. \cite{cheng2014cloud} find the most suitable pixel for the corrupted area with a spatio-temporal Markov random field (MRF) model. \cite{zhang2020thick} fill cloudy regions by a spatio-temporal patch group deep learning framework. \cite{benabdelkader2007cloud} improve the contextual reconstruction process with spatio-spectral information.

SAR images are widely used in cloud removal because of their advantages, such as passing through clouds and smoke. They can be obtained 24 hours a day and regardless of the weather conditions. However, disadvantages such as speckle noise, lack of color information, and geometry distortion and shadows cause experts unable to distinguish between different areas. Using SAR-to-optical translated images is a way to solve this problem \citep{wang2019sar}.

Deep learning is a powerful tool to find hidden structures in large data sets by the self feature extracting \citep{lecun2015deep}. In recent years, it has improved the performance of remote sensing image analysis tasks such as scene classification, object detection, and land use and land cover (LULC) classification \citep{ma2019deep}. Generative Adversarial Networks (GANs) \citep{goodfellow2014generative} have seen a massive rise in popularity among deep learning methods because of their results, especially in the image-to-image translation tasks such as pix2pix \citep{isola2017image} and Cycle-Consistent Adversarial Networks (CycleGAN) \citep{zhu2017unpaired}. To generate fake images similar to real images, \cite{mirza2014conditional} introduced conditional generative adversarial network (cGAN) by assigning latent space elements to particular data distribution. \cite{bermudez2019synthesis} used cGAN for SAR-to-optical translation with SAR/optical multitemporal data. \cite{singh2018cloud} trained a CycleGAN to remove clouds.
 
Based on the aforementioned studies, we introduce a model to cloud removal in remote sensing images using GANs.
The main contributions of this work are summarized as follows:
\begin{enumerate}[1.]
\item Our model includes two GANs. We train the former to discover the intricate structures in SAR images with large datasets. The latter is used to remove clouds and retains the quality of cloud-free areas.
\item We propose DRIBs for generators based on dilated convolutions to increase receptive view (global view) and prevent missing information. 
\item We train and test our model with the SEN1-2 dataset and compare it with state-of-the-art deep learning-based cloud removal methods. The results show the superiority of our model in both SAR-to-optical and cloud removal tasks.
\end{enumerate}

The rest of this paper is structured as follows. In Section 2, the proposed method is explained. The comparison results are presented in Section 3. Also, we show the influence of the SAR-to-optical step, DRIBs, and SSIM loss function in the ablation study. Section 4 provides a discussion on the results, and a conclusion is given in Section 5.


\section{Methods}
\begin{figure*}[!t]
\begin{center}
\includegraphics[width=1\linewidth]{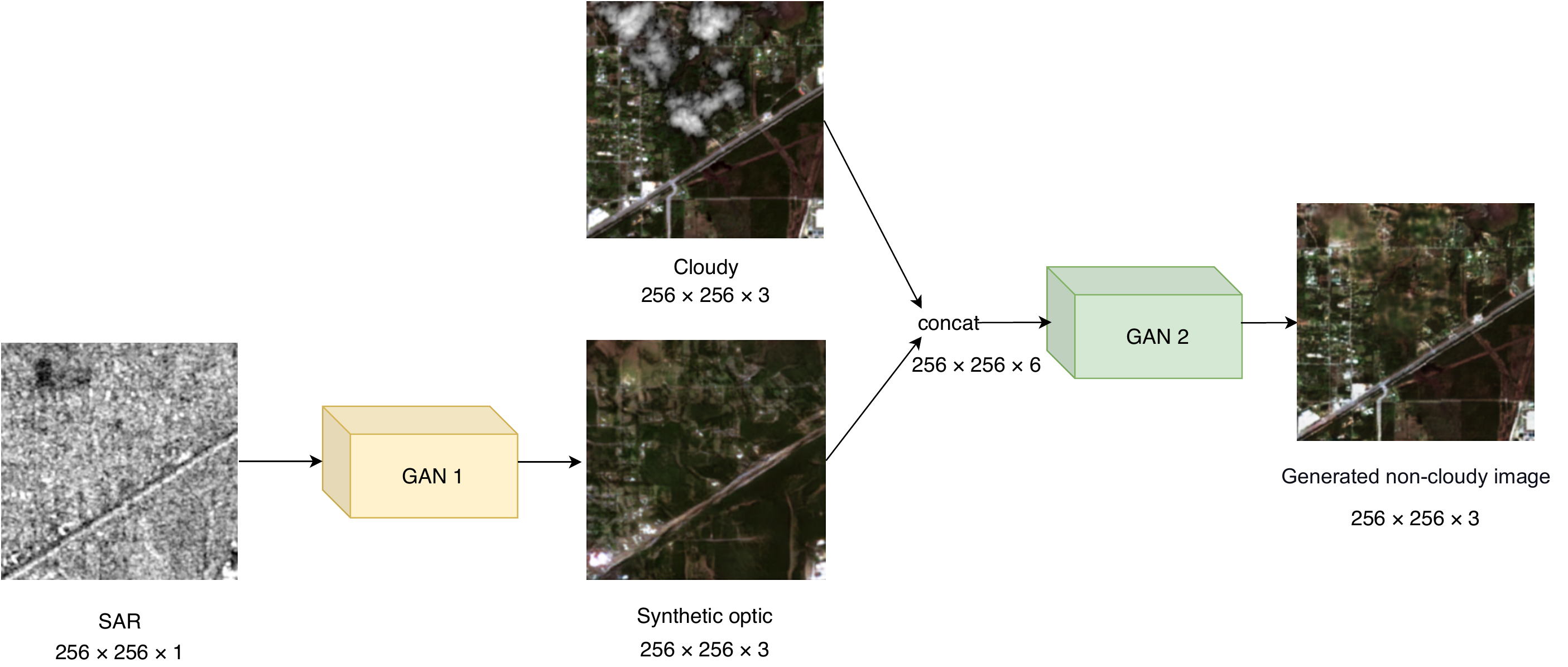}
\end{center}
   \caption{Overview of the model. The tasks of GAN1 and GAN2 include SAR-to-optical translation and cloud removal, respectively.}
\label{fig1}
\end{figure*}

Unlike thin clouds, thick clouds completely block surface information, and we have no information about the areas below the cloud. SAR images help us as auxiliary images for restoring cloudy regions because of their ability to cross the cloud. Translating SAR images to synthetic optic images instead of using SAR images directly to remove the clouds can improve the quality of output images \citep{wang2019sar}. The success of GANs in image translation has led us to use them in our method. We use two GANs for the following reasons: (1) The cost of obtaining a large-scale dataset of SAR and optical pair images is less than acquiring co-registered optical cloudy, cloud-free, and SAR images. (2) The training of SAR-to-optical image translation needs many images. As shown in Fig. \ref{fig1}, our method consists of two parts: translation of SAR  to optical images and removing clouds. This separation of tasks improves the training phase of SAR-to-optical translation GAN with many available images, and the limitation of the number of co-registered cloudy and cloud-free images does not affect it.

\subsection{Network Architecture}
\begin{figure*}[!t]
\begin{center}
\includegraphics[width=1\linewidth]{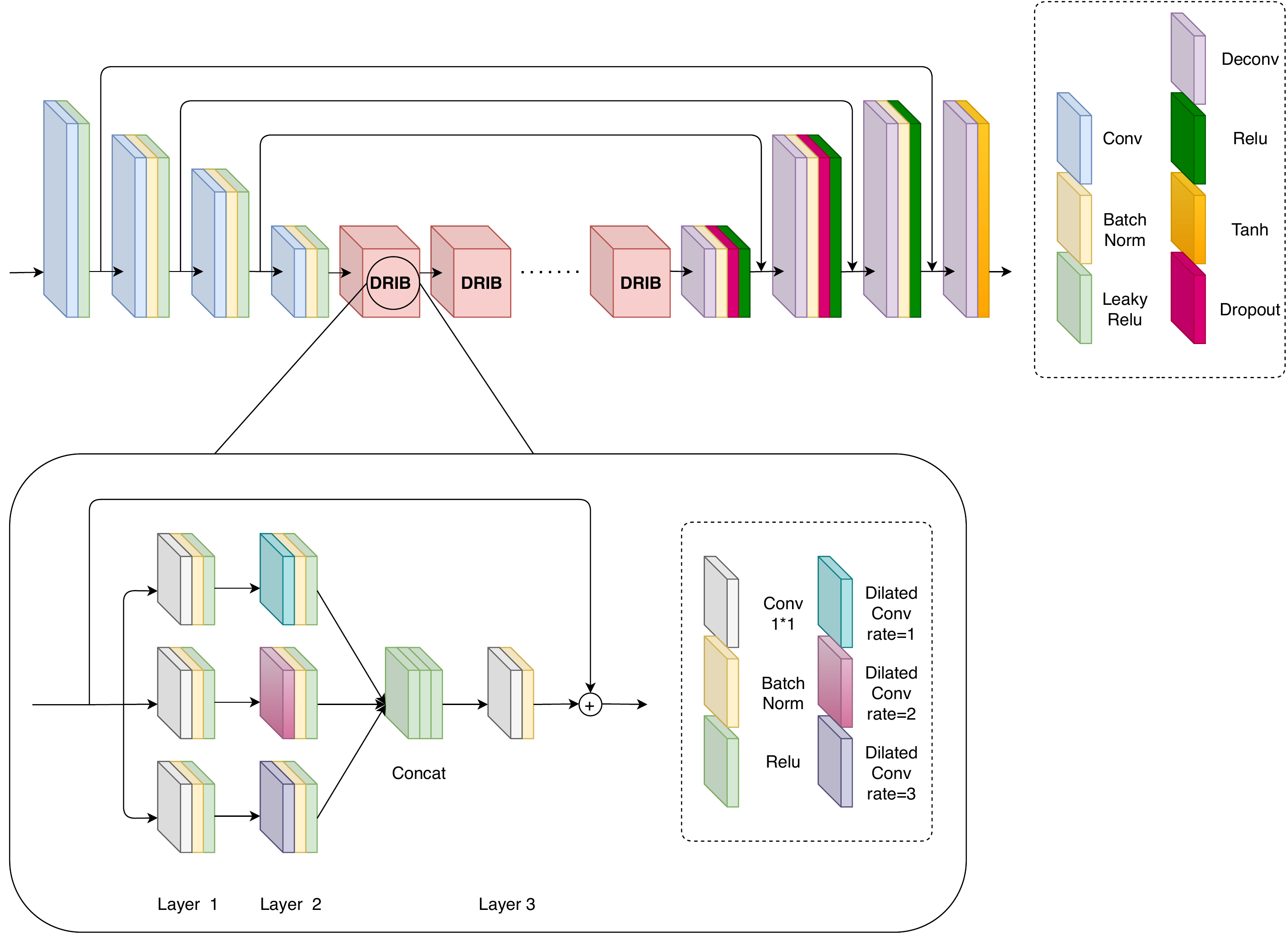}
\end{center}
   \caption{The architecture of the generator consists of an encoder, DRIBs, and decoder. A DRIB structure is magnified.}
\label{fig2}
\end{figure*}

For both GANs, we propose the encoder-DRIBs-decoder structure in the generator, as shown in Fig. \ref{fig2}. We adopt symmetrical concatenations between encoder and decoder for giving local information to upsampled outputs \citep{ronneberger2015u}. The details of generator and discriminator networks are presented in Table \ref{table1}. Table \ref{table2} shows DRIB architecture, which we used 8 of it as a bottleneck.

\begin{table}[H]
\caption{Details of discriminator and generator except DRIBs. Information flow is from top to bottom within components. Acronyms: C=convolution, B=batch normalization, L= leaky relu, D=deconvolution, R=relu, T=tanh. The numbers in parentheses indicate the number of filters, filter size, and stride of the convolution filters, respectively.}
\begin{center}
\begin{tabular}{lll}
\hline
Encoder                                                                                         & Decoder                                                                                         & Discriminator                                                                                               \\ \hline
\begin{tabular}[c]{@{}l@{}}CL(64,4,2)\\ CBL(128,4,2)\\ CBL(256,4,2)\\ CBL(512,4,1)\end{tabular} & \begin{tabular}[c]{@{}l@{}}CBDR(256,4,1)\\ CBDR(128,4,2)\\ CBR(64,4,2)\\ CT(3,4,2)\end{tabular} & \begin{tabular}[c]{@{}l@{}}CBL(64,4,2)\\ CBL(128,4,2)\\ CBL(256,4,2)\\ CBL(512,4,2)\\ C(1,3,1)\end{tabular} \\ \hline
\end{tabular}
\end{center}
\label{table1}
\end{table}


\begin{table}[H]
\caption{Details of DRIBs in corresponding layers of Fig. \ref{fig2}. Acronyms: C=convolution, B=batch normalization, D=deconvolution, R=relu. The numbers in parentheses indicate the number of filters, filter size, stride, and dilation rate of the convolution filters, respectively.}
\begin{center}
\begin{tabular}{lll}
\hline
Layer 1     & Layer 2        & Layer 3     \\ \hline
CBR(256,1,1,1) & DBR(256,3,1,1)  &              \\ \cline{1-2}
CBR(256,1,1,1) & DBR(256,3,1,2)  & CB(512,1,1,1)  \\ \cline{1-2}
CBR(256,1,1,1) & DBR(256,3,1,3)  &              \\ \hline
\end{tabular}
\end{center}
\label{table2}
\end{table} 

\begin{figure}[!t]
\begin{center}
\includegraphics[width=\linewidth]{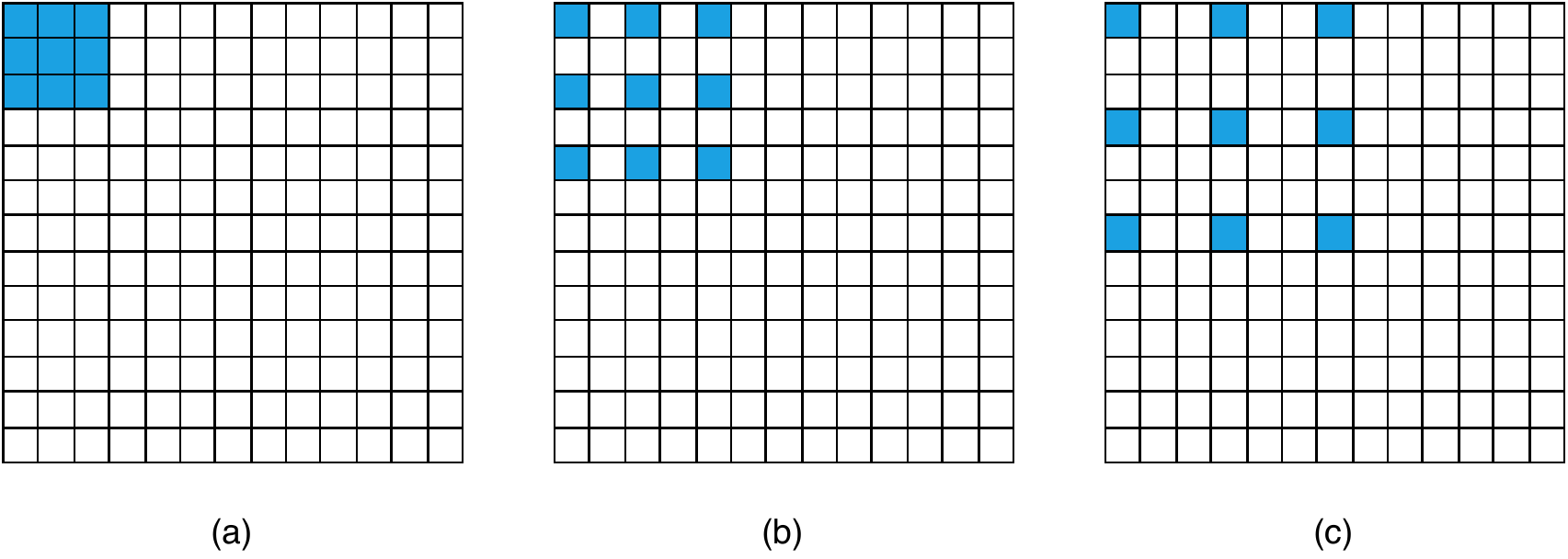}
\end{center}
   \caption{3 × 3 Dilated convolutions with different dilation rate.  (a) Dilated convolution, r=1. (b) Dilated convolution, r=2. (c) Dilated convolution, r=3. Dilated convolutions keep the spatial resolution by expanding the receptive field.}
\label{fig4}
\end{figure}

The main features of DRIB architecture are as follows:
\begin{enumerate}[1.]
\item \textbf{Dilated convolutions:} Because of the loss of spatial information by subsampling \citep{yu2017dilated}, we reduce the number of stride-2 convolutions by replacing dilated convolutions. Dilated convolutions prevent losing resolution by expanding the receptive field \citep{yu2015multi}, as shown in Fig. \ref{fig4}.
\item \textbf{Residual connections:} Learning becomes more difficult as networks deepen. The cause is the problem of vanishing/exploding gradients, which prevent the weights from changing their values. The residual connection is a way to tackle this problem \citep{he2016deep}.
\item \textbf{Inception modules:} We applied inception modules to process visual information at different scales \citep{szegedy2015going}. Also, it helped to reduce the depth of the network and allows for more efficient computations. Residual Inception Blocks were emerged by combining the inceptions modules and residual connections. This combination caused accelerating the training of inception networks \citep{szegedy2016inception}.
\end{enumerate}
Using these features for DRIBs halved the number of our generator parameters than U-net \citep{ronneberger2015u} but raised performance.

\subsection{Objective}
We used the objective of conditional GAN to both generator and discriminator networks, which can be expressed as:
\begin{equation}
\begin{split}
L_{cGAN}=E_{x,y\sim p_{data}(x,y)}\lbrack log(D(x,y))\rbrack 
+E_{x,z\sim p_{data}(x,z)}\lbrack log(1-D(x,G(x,z)))\rbrack
\end{split}
\end{equation}
where D represents the discriminator network, G represents the generator network, and D tries to maximize the objective against an adversarial G that tries to minimize it.

To reduce blurring and bring the output image closer to the target image, we used L1 loss function as follows:
\begin{equation}
L_{1}= \vert \vert G(x)-y\vert \vert _{1} 
\end{equation}
where y represents the target image.

SSIM measures the similarity between two images and assesses quality based on the degradation of structural information \citep{wang2004image}. It is calculated between two images x and y as follows:
\begin{equation}
SSIM_{(x,y)}= \frac{(2\mu_{x}\mu_{y}+C_{1})}{(\mu^2_{x}+\mu^2_{y}+C_{1})}\cdot\frac{(2\sigma_{xy}+C_{2})}{(\sigma^2_{x}+\sigma^2_{y}+C_{2})}
\label{eq:3}
\end{equation}
where $\mu$, $\sigma^2$, and $\sigma_{xy}$ are the average, variance, and covariance, respectively. $C_{1}$ and $C_{2}$ are variables to stabilize the division with weak denominator.

The SSIM loss function in patch P can be written as:
\begin{equation}
L_{ssim}(P)= \frac{1}{N} \sum(1-SSIM_{p})
\end{equation}
where p is the center pixel of patch P. The size of the patch and Gaussian filter is 11 × 11.

We added SSIM loss function to the pix2pix loss function to learn to produce visually delightful images \citep{zhao2016loss}. The final objective of our generators is computed via:
\begin{equation}
L_{G}= L_{cGAN}+ \lambda_{1} L_{1}+ \lambda_{2} L_{ssim}
\end{equation}
where $\lambda_{1} = 100$, $\lambda_{2} = 100$.
\subsection{Training and testing}
The GANs are trained with the SEN1-2 dataset \citep{schmitt2018sen1} consisting of co-registered Sentinel-1 SAR and Sentinel-2 optical image patches with 256 × 256 px. Only VV and RGB bands were used for the dataset with a spatial sampling distance of 10m. All pixel values of images are normalized to [-1, 1]. We trained SAR-to-optical translation GAN with 18712 pair images and tested it with 4678. 

\begin{figure*}[!t]
\begin{center}
\includegraphics[width=0.7\linewidth]{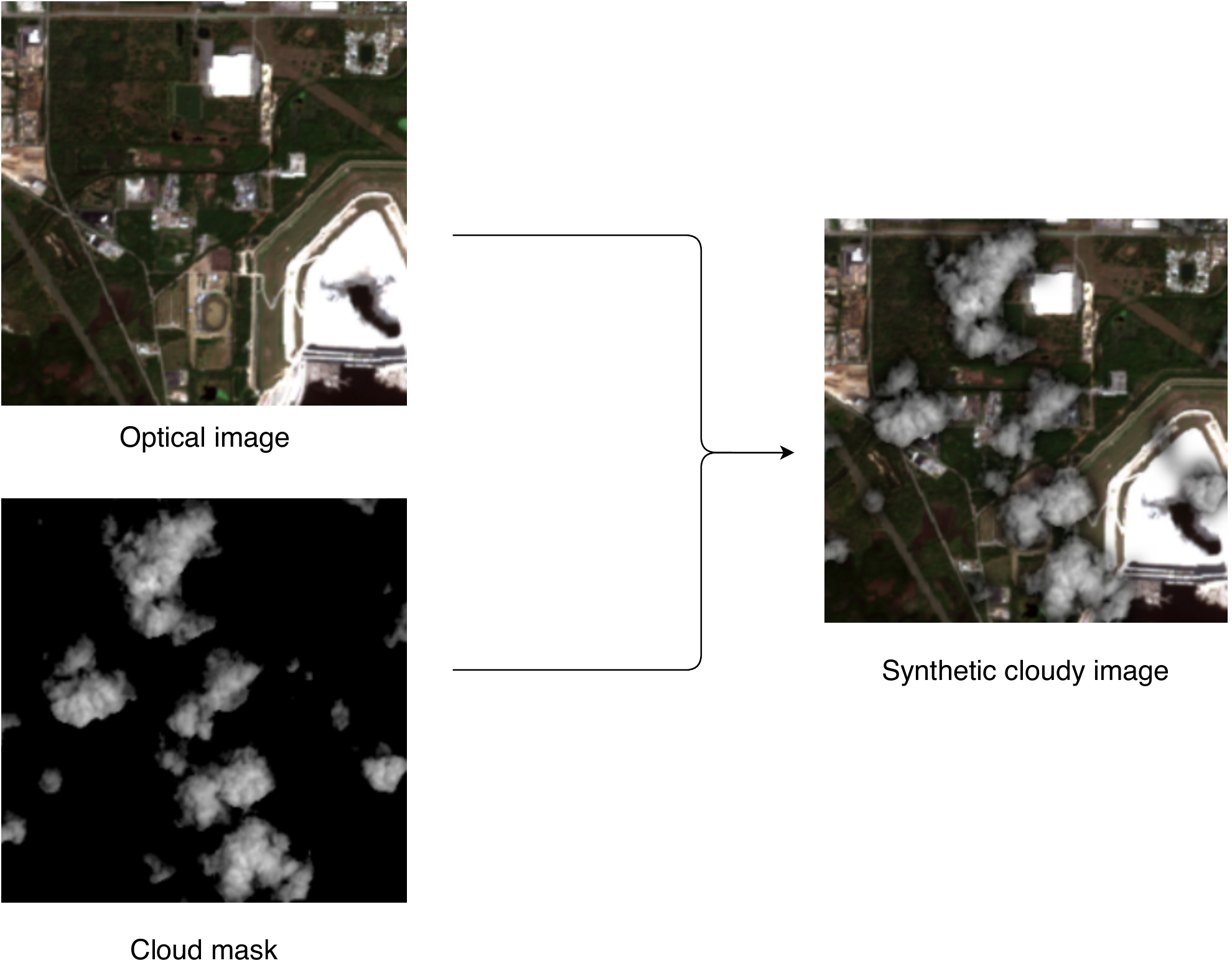}
\end{center}
   \caption{Alpha blending process}
\label{fig10}
\end{figure*}

\begin{figure*}[!t]
\begin{center}
\includegraphics[width=0.7\linewidth]{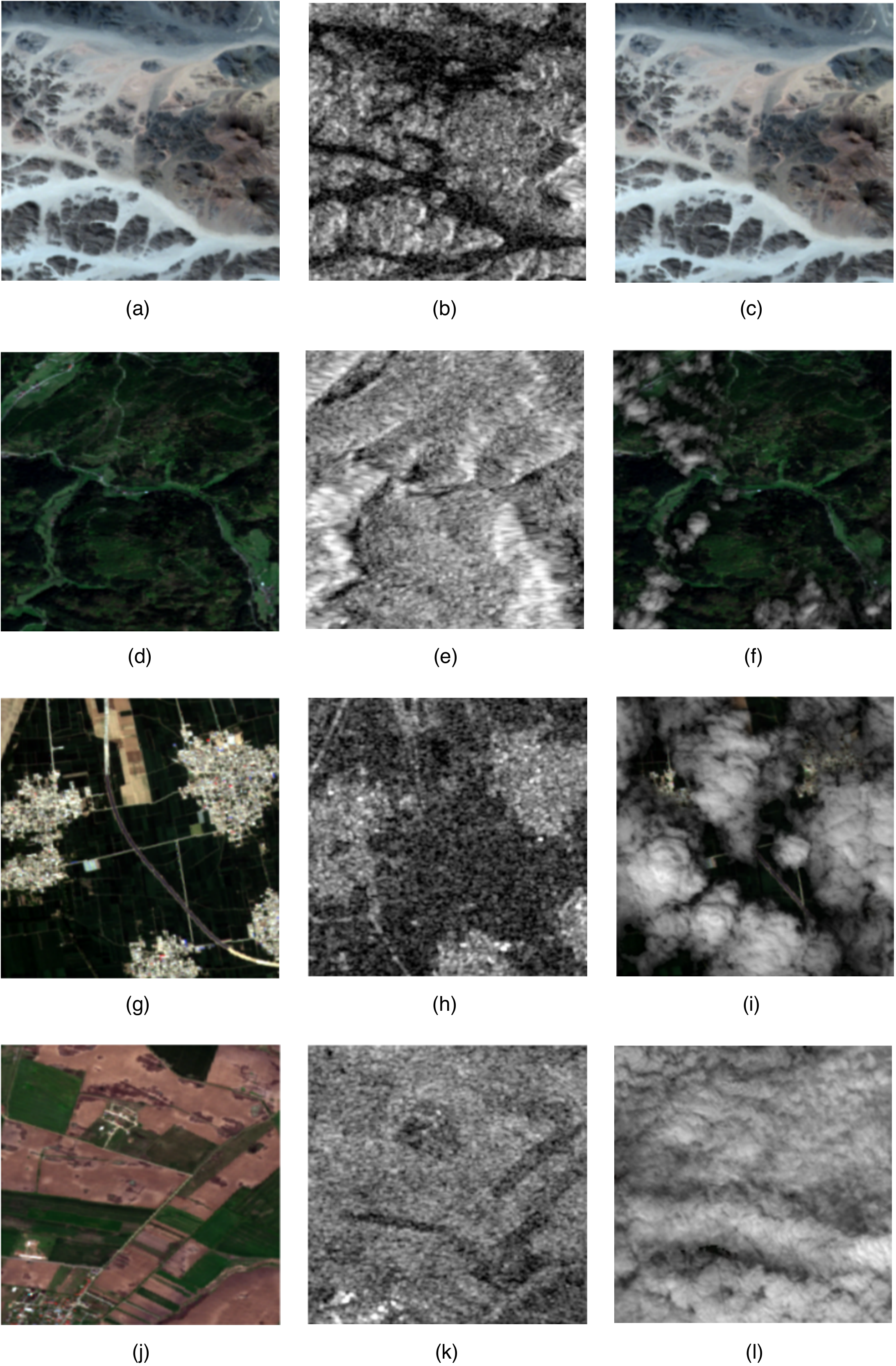}
\end{center}
   \caption{Examples of dataset; left: optical image, center: SAR image, right: cloudy image.}
\label{fig0}
\end{figure*}

For cloudy images, we used thick synthetic clouds that had been clipped from real cloudy images. Then cloud masks were added to optical images by alpha blending \citep{porter1984compositing}, as shown in Fig. \ref{fig10}. We eliminate background information from cloudy regions by this approach. Fig. \ref{fig0} shows four examples of our dataset with different cloud percentages. We trained cloud removal GAN with 2000 pair images and tested it with 500. 

\section{Experimental results}
In this section, we will first explain the details of the training. Then we demonstrate the evaluation metrics used, and we report the results of our comparisons based on them. Finally, we investigate the effect of DRIBs and SSIM loss function.
\subsection{Implementation details}
For both GANs, we set the following hyper-parameters: learning rate $\alpha=2^{-4}$, batch size=32, dropout=0.5. We initialized the weights of the networks with normal distribution $\mathcal{N}(\mu=0,\sigma^2=0.02)$. We used Adam optimizer with momentum parameters $\beta_{1}=0.5$ and $\beta_{2}=0.999$.
The size of the inputs of the discriminator in SAR-to-optical translation GAN is set to 256 × 256 × 1 and 256 × 256 × 3. These sizes are 256 × 256 × 6 and 256 × 256 × 3 in cloud removal GAN.

\subsection{Metrics}
We evaluated our model in both SAR-to-optical translation and cloud removal parts with SSIM eq. \ref{eq:3} and peak signal to noise ratio (PSNR){} as given by:
\begin{equation}
PSNR = 20 \cdot \log_{10} (\frac{MAX_{I}}{\sqrt{MSE} })
\end{equation}
where $MAX_{I}$ represents the maximum possible pixel value of the image.

\subsection{Performance comparison}
This section compares the proposed method with state-of-the-art approaches in SAR-to-optical translation and cloud removal separately.
\subsubsection{SAR-to-optical image translation}
\begin{figure*}[!t]
\begin{center}
\includegraphics[width=1\linewidth]{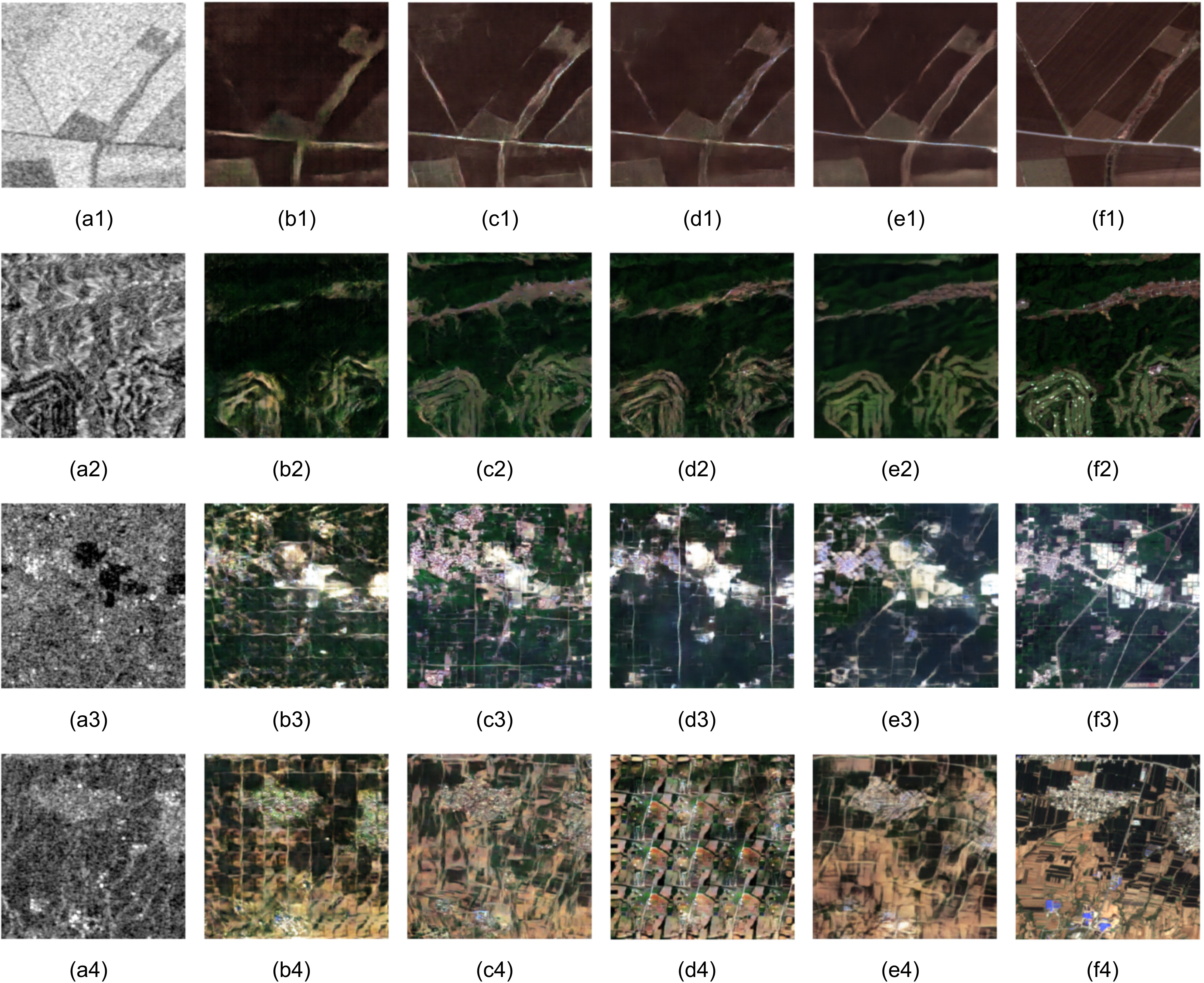}
\end{center}
   \caption{Example results of SAR to opt translation GAN. Column 1 shows SAR images, Column 2 results of \citep{bermudez2018sar} model, Column 3 results of \citep{enomoto2018image} model, Column 4 results of  \citep{li2020sar} model, Column 5 results of our model, and Column 6 target images.}
\label{fig5}
\end{figure*}

We compared our SAR-to-optical translation GAN with three deep learning-based models. \citep{bermudez2018sar,enomoto2018image,li2020sar} used GANs with different architectures to translate SAR-to-optical images. The samples of generator outputs are presented in Fig. \ref{fig5}. Table \ref{table3} shows corresponding quantitative results.

\begin{table}[H]
\caption{SAR to Opt translation performance comparison.}
\begin{center}
\begin{tabular}{lll}
\hline
Method                                     & PSNR    & SSIM   \\ \hline
\citep{bermudez2018sar}       & 17.1878 & 0.2981 \\ \hline
\citep{enomoto2018image}           & 18.6526 & 0.3606 \\ \hline
\citep{li2020sar} & 19.0713 & 0.4196 \\ \hline
Ours                                & \textbf{19.4495} & \textbf{0.4453} \\ \hline
\end{tabular}
\end{center}
\label{table3}
\end{table}

\subsubsection{Cloud removal}
\begin{figure*}[!t]
\begin{center}
\includegraphics[width=1\linewidth]{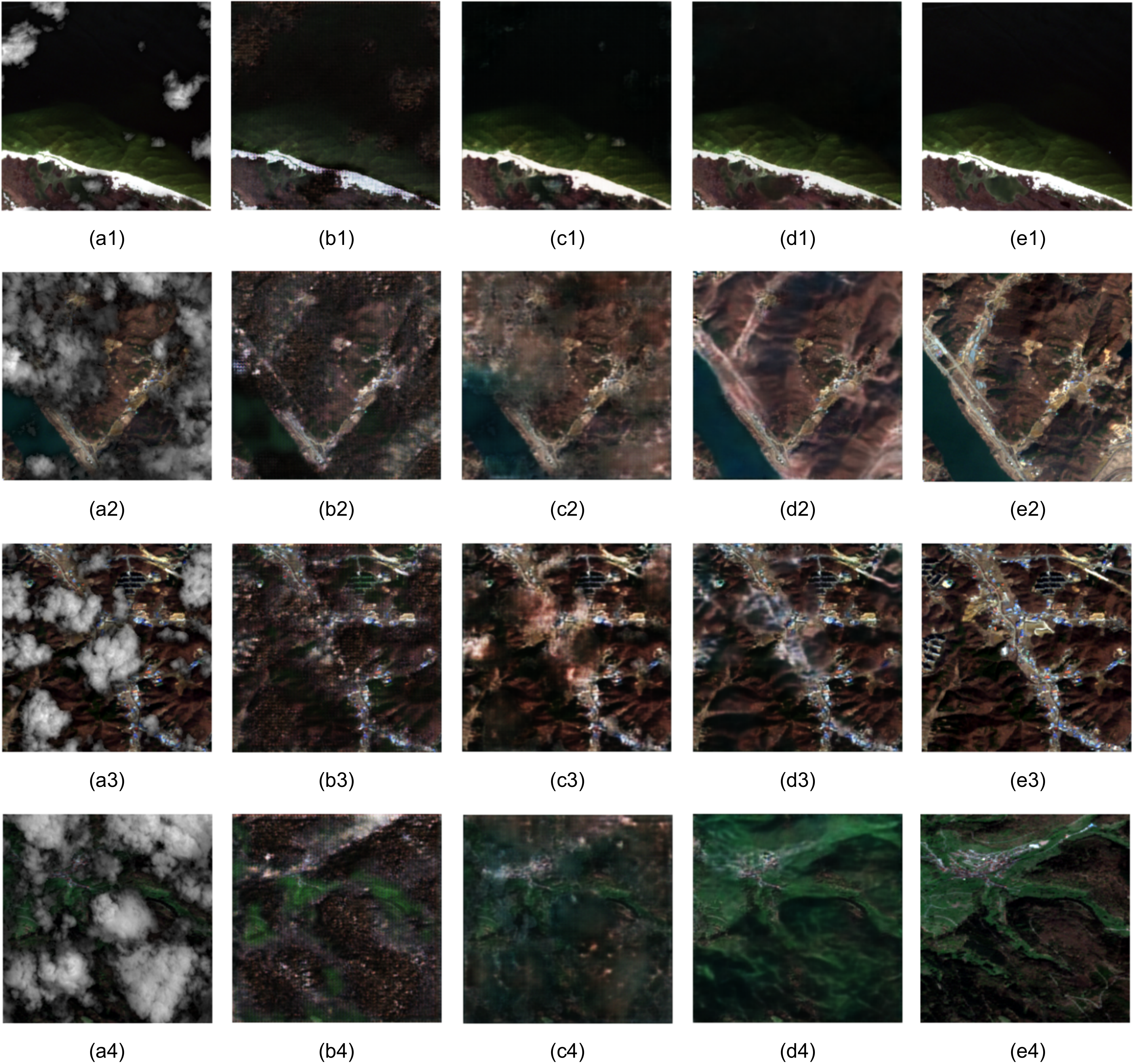}
\end{center}
   \caption{Example results of cloud removal GAN. Column 1 shows cloudy images, Column 2 results of \citep{ebel2020multisensor} model, Column 3 results of \citep{grohnfeldt2018conditional} model, Column 4 results of our model, and Column 5 target images.}
\label{fig6}
\end{figure*}

To compare our cloud removal GAN, we use two GAN based methods, which used SAR and optical data fusion. \citep{grohnfeldt2018conditional} presented a cGAN to remove clouds with SAR. \cite{ebel2020multisensor} removed clouds using CycleGAN and auxiliary loss term. Fig. \ref{fig6} presents cloudy images and corresponding cloud-free images generated by different methods. The PSNR and SSIM values are listed in Table \ref{table4}. We measured PSNR for cloudy and non-cloudy areas separately to be able to assess the cloud removal power of models in cloudy areas accurately.

\begin{table}[H]
\caption{Cloud removal performance comparison.}
\begin{center}
\resizebox{\linewidth}{!}{%
\begin{tabular}{lllll}
\hline
Method               & PSNR    & SSIM   & \begin{tabular}[c]{@{}l@{}}PSNR (cloudy areas)\end{tabular} & \begin{tabular}[c]{@{}l@{}}PSNR (non-cloudy areas)\end{tabular} \\ \hline
\citep{ebel2020multisensor}  & 17.8966 & 0.4863 & 14.1949                                                        & 21.9991                                                            \\ \hline
\citep{grohnfeldt2018conditional} & 19.2701 & 0.5535 & 15.9657                                                        & 22.7272                                                            \\ \hline
Ours          & \textbf{23.0941} & \textbf{0.6909} & \textbf{19.9044}                                                        & \textbf{25.9061} \\ \hline
\end{tabular}}
\end{center}
\label{table4}
\end{table}

\subsection{Ablation study}
\subsubsection{Effect of SAR-to-optical image translation step and DRIBs}

To prove the influence of the SAR-to-optical translation and DRIBs, we examined our model with different components for generators, as shown in Table \ref{table5}. In the second row of Table \ref{table5}, the SAR-to-optical translation is not used, and a cloud image with a SAR image are given to the input of the cloud removal generator. In the rest of the rows, we replaced DRIBs with U-net. The results show the effect of our model parts on improving the quality of output images.

\begin{table}[H]
\caption{Performance of our method when using different architectures for the generator.}
\begin{center}
\resizebox{\linewidth}{!}{%
\begin{tabular}{llllll}
\hline
Cloud Removal & SAR2OPT & PSNR    & SSIM   & \begin{tabular}[c]{@{}l@{}}PSNR (cloudy areas)\end{tabular} & \begin{tabular}[c]{@{}l@{}}PSNR (non-cloudy areas)\end{tabular} \\ \hline
DRIBs           & \_      & 16.5430 & 0.4515 & 15.4963                                                        & 17.9218                                                            \\ \hline
U-NET          & U-NET    & 21.3734 & 0.6150 & 19.7257                                                        & 22.80.87                                                           \\ \hline
DRIBs           & U-NET    & 21.8911 & 0.6363 & 19.7043                                                        & 23.8589                                                            \\ \hline
U-NET          & DRIBs     & 22.6575 & 0.6835 & 19.8227                                                        & 25.1819                                                            \\ \hline
DRIBs           & DRIBs     & \textbf{23.0941} & \textbf{0.6909} & \textbf{19.9044}                                                        & \textbf{25.9061}                                                            \\ \hline
\end{tabular}}
\end{center}
\label{table5}
\end{table}

\subsubsection{Effect of SSIM loss function}
In order to show the impact of the SSIM loss function, we abandoned it in different modes. Table \ref{table6} presents results of our investigation. PSNR and SSIM values increased by adding SSIM loss function.

\begin{table}[H]
\caption{Performance of our method when using different loss functions.}
\begin{center}
\resizebox{\linewidth}{!}{%
\begin{tabular}{llllll}
\hline
Cloud Removal & SAR2OPT & PSNR    & SSIM   & \begin{tabular}[c]{@{}l@{}}PSNR (cloudy areas)\end{tabular} & \begin{tabular}[c]{@{}l@{}}PSNR (non-cloudy areas)\end{tabular} \\ \hline
L1            & L1      & 20.4185 & 0.5304 & 18.6577                                                        & 21.7619                                                            \\ \hline
SSIM+L1       & L1      & 21.3423 & 0.6063 & 19.0759                                                        & 23.0808                                                            \\ \hline
SSIM          & SSIM    & 21.8548 & 0.6694 & 18.5796                                                        & 24.7527                                                            \\ \hline
L1            & SSIM+L1 & 22.0853 & 0.6309 & 19.4336                                                        & 24.3254                                                            \\ \hline
SSIM+L1       & SSIM+L1 & \textbf{23.0941} & \textbf{0.6909} & \textbf{19.9044}                                                        & \textbf{25.9061}                                                            \\ \hline
\end{tabular}}
\end{center}
\label{table6}
\end{table}


\section{Discussion}
As the experimental results in Section 3 show, our model performs better than state-of-the-art methods. Tables \ref{table3} and \ref{table4} show that the proposed model has the highest PSNR and SSIM values in both SAR-to-optical translation and cloud removal parts. According to the PSNR values in cloudy and non-cloudy areas, it can be concluded that our model can remove the cloud very well and also keep the non-cloudy areas unchanged. This performance was due to using two GANs, changing the network structure, and adding the SSIM loss function. As explained in Section 2 about DRIB architecture, we expected better performance than classical U-net. Table \ref{table5} confirms DRIBs performance. Our results also showed that adding the SSIM loss function to the classical pix2pix loss function increased the PSNR and SSIM in the output images, as shown in Table \ref{table6}.


\section{Conclusions}

In this paper, we proposed a model to remove clouds using two GANs and SAR images. Compared to previous GAN-based methods, our method has improvements in the two parts of the generator network structure and the loss function. The experimental results on the SEN1-2 dataset show the superiority of our model over the other state-of-the-art methods in both SAR-to-optical translation and cloud removal parts.
 
Developing a cloud removal model using real cloudy images and utilizing its generated images for classification tasks can be considered further. Providing a solution to remove the cloud from unpaired images can be another issue for future work.

\section*{Acknowledgment}

This work was supported by the Center for Space Research of Iran.

\end{document}